\documentclass{article}
\usepackage{amsmath,amssymb,amsbsy,amsfonts,latexsym,amsopn,amstext,amsxtra,euscript,amscd,color, amsthm}

\newtheorem{thm}{Theorem}
\newtheorem{lem}{Lemma}
\newtheorem{cor}{Corollary}

\newtheorem{conj}{Conjecture}
\newtheorem{exmp}{Example}
\newtheorem{defn}{Definition}

\newcommand{\K}{\mathbb{K}}
\newcommand{\Q}{\mathbb{Q}}
\newcommand{\C}{\mathbb{C}}
\newcommand{\car}{\mathrm{char}}
\newcommand{\F}{\mathbb{F}}

\begin{document}

\title{On decomposition of tame polynomials and rational functions}

\author{\small Jaime Gutierrez\\\small Dpto. de Matem\'aticas, Estad\'{\i}stica y Computaci\'on, Univ. de Cantabria\\\small E--39071 Santander, Spain\\\small jaime.gutierrez@unican.es \and \small David Sevilla\\\small Department of Computer Science, Concordia University\\\small Montreal H3G 1M8, QC, Canada\\\small dsevilla@cs.concordia.ca}

\date{}

\maketitle

\begin{abstract}
In this paper we present algorithmic considerations and theoretical results about the relation
between the orders of certain groups associated to the components of a polynomial and the order of
the group that corresponds to the polynomial, proving it for arbitrary tame polynomials, and
considering the case of rational functions.
\end{abstract}

\section{Introduction}

The general functional decomposition problem can be stated as follows: given $f$ in a class of
functions, we want to represent $f$ as a composition of two ``simpler" functions $g$ and $h$ in
the same class, i.e. $f=g\circ h=g(h)$. Although not every function can be decomposed in this
manner, when such a decomposition does exist many problems become significantly simpler. \par

Univariate polynomial decomposition has applications in computer science, computational algebra,
and robotics. In fact, computer algebra systems such as {\sc Axiom}, {\sc Maple}, {\sc
Mathematica}, and {\sc Reduce} support polynomial decomposition for univariate polynomials. For
some time, this problem was considered computationally hard: the security of a cryptographic
protocol was based on its hardness, see~\cite{Cade}. A polynomial time algorithm is given in
\cite{KL89}, requiring $O(ns\log r)$ or $O(n^2)$ field operations, where $n=\deg f$, $r=\deg g$,
and $s=\deg h$. It works over any commutative ring in the \emph{tame case}, that is, when the ring
contains a multiplicative inverse of $r$, and assumes that the polynomials involved are monic.
Independently, \cite{GRR} presented a similar algorithm, running in time $O(n^2)$ sequentially and
$O(n\log^2n)$ in parallel. Several papers have been published on different extensions and
variations of this problem; see for instance~\cite{Gat90},\cite{Gat91}, \cite{CFM}, \cite{GGR} and
\cite{Gut91}.

In \cite{Zip} a polynomial time algorithm to decompose a univariate rational function over any
field is presented with efficient polynomial factorization. The paper \cite{AGR95} presented two
exponential-time algorithms to decompose rational functions, which are quite efficient in
practice. They have been implemented in the {\sc Maple} package {CADECOM}, which is designed for
performing computations in rational function fields; see~\cite{GR00}.

In this paper we will focus on certain structural properties of decomposition of polynomials and
rational functions in one variable. Namely, for each polynomial or rational function $f$ in one
variable, we can consider the group of transformations of the form
\[z\ \mapsto\ \frac{az+b}{cz+d}\qquad \mathrm{such\ that}\qquad f(z)=f\left(\frac{az+b}{cz+d}\right).\]
The relation between the degree of a rational function and the order of its corresponding group
can provide valuable information about the structure of the different decompositions of the
function. In particular, the following result appears in \cite{BN00}:

\begin{thm}[\cite{BN00}]
Let $p_1,\ldots,p_m\in\C[x]$ be non-constant and $k_1,\ldots,k_m,k$ be the orders of the groups
$G(p_1),\ldots,G(p_m),G(p_1\circ\cdots\circ p_m)$. Then $k$ divides $k_1\cdots k_m$.
\end{thm}

One of our goals is to generalize this result to a wide class of polynomials, namely the
\emph{tame polynomials}, and also consider other generalizations, like the case of rational
functions. We think that it can be used to obtain better algorithms for decomposing non tame
polynomials, see \cite{Gat91}.

\section{Polynomial and rational decomposition}

Our starting point is the decomposition of polynomials and rational functions in one variable.
First we will define the basic concepts of this topic in full generality.

\begin{defn}
Let $\K$ be any field, $x$ a transcendental over $\K$ and $\K(x)$ the field of rational functions
in the variable $x$ with coefficients in $\K$. In the set $T=\K(x)\setminus\K$ we define the
binary operation of \emph{composition} as
\[g(x)\circ h(x)=g(h(x))=g(h).\]

We have that $(T,\circ)$ is a semigroup, the element $x$ being its neutral element.

If $f=g\circ h$, we call this a \emph{decomposition} of $f$ and say that $g$ is a \emph{component
on the left} of $f$ and $h$ is a \emph{component on the right} of $f$. We call a decomposition
\emph{trivial} if any of the components is a unit with respect to decomposition; the units in
$(T,\circ)$ are precisely the elements of the form
\[\frac{ax+b}{cx+d},\quad a,b,c,d\in\K,\quad ad-bc\neq 0.\]

Given two decompositions $f=g_1\circ h_1=g_2\circ h_2$ of a rational function, we call them
\emph{equivalent} if there exists a unit $u$ such that
\[h_1=u\circ h_2,\ g_1=g_2\circ u^{-1},\]
where the inverse is taken with respect to composition.

Given $f\in T$, we say that it is \emph{indecomposable} if it is not a unit and all its
decompositions are trivial.

We define a \emph{complete} decomposition of $f\in\K(x)$ to be $f=g_1\circ\cdots\circ g_r$ where
$g_i$ is indecomposable. The notion of equivalent complete decompositions is straightforward from
the previous concepts.
\end{defn}

\begin{defn}
Given a non--constant rational function $f(x)\in\K(x)$ where $f(x)=f_N(x)/f_D(x)$ with
$f_N,f_D\in\K[x]$ and $(f_N,f_D)=1$, we define the \emph{degree} of $f$ as
\[\deg\,f=\max\{\deg\,f_N,\ \deg\,f_D\}.\]

We also define $\deg\,a=0$ when $a\in\K$.
\end{defn}

\noindent\textbf{Remark.} From now on, we will use the previous notation when we refer to the
numerator and denominator of a rational function. Unless explicitly stated, we will take the
numerator to be monic, even though multiplication by constants will not be relevant.

Now we introduce some basic results about univariate decomposition, see \cite{AGR95} for more
details.

\begin{lem} $ $
\begin{description}
   \item[(i)] For every $f\in T$, $\deg\,f=[\K(x):\K(f)]$.
   \item[(ii)] $\deg\,(g\circ h)=\deg\,g\cdot\deg\,h$.
   \item[(iii)] $f(x)$ is a unit with respect to composition if
and only if $\deg\,f=1$, that is,
$f(x)=\displaystyle\frac{ax+b}{cx+d}$ with $a,b,c,d\in\K$ and
$ad-bc\not=0$.
   \item[(iv)] Every non--constant element of $\K(x)$ is cancellable on
the right with respect to composition. In other words, if
$f(x),h(x)\in T$ are such that $f(x)=g(h(x))$ then $g(x)$ is
uniquely determined by $f(x)$ and $h(x)$.
\end{description}
\end{lem}

Now we relate decomposition and Field Theory by means of the following extended version of
L\"uroth's theorem.

\begin{thm}\label{teo-lurext}
Let $\K(\mathbf{x})=\K(x_1,\dots,x_n)$ be the field of rational functions in the variables
$\mathbf{x}=(x_1,\ldots,x_n)$ over an arbitrary field $\K$. If $\F$ is a field of transcendence
degree 1 over $\K$ with $\K\subset \F\subset \K(\mathbf{x}) $, then there exists
$f\in\K(\mathbf{x})$ such that $\F = \K(f)$. Moreover, if $\F$ contains a non--constant polynomial
over $\K$, then there exists a polynomial $f\in \K[\mathbf{x}]$ such that $\F = \K(f)$.
\end{thm}

\begin{proof}
For a proof we refer to \cite{Sch82}, Theorems 3 and 4, and \cite{Nag93}. Constructive proofs can
be found in \cite{Net1885} for $n=1$, and in \cite{GRS02} for arbitrary $n$.
\end{proof}

Let $f=g\circ h$. Then $f\in\K(h)$, thus $\K(f)\subset\K(h)$. Also, $\K(f)=\K(h)$ if and only if
$f=u\circ h$ for some unit $u$. This provides the following bijection between the decompositions
of a rational function $f$ and the intermediate fields in the extension $\K(f)\subset\K(x)$.

\begin{thm}\label{equiv-dec-field}
Let $f\in\K(x)$. In the set of decompositions of $f$ we have an equivalence relation given by the
definition of equivalent decompositions, and we denote as $[(g,h)]$ the class of the decomposition
$f=g\circ h$. Then we have the bijection:
\[\begin{array}{ccc}
\{\,[(g,h)]:f=g(h)\,\} & \longleftrightarrow & \{\,\F:\K(f)\subset\F\subset\K(x)\,\} \\
\left[(g,h)\right] & \longleftrightarrow & \F=\K(h).
\end{array}\]
\end{thm}

Of special interest is the case of $f$ being a polynomial. The following corollary to the second
part of Theorem~\ref{teo-lurext} shows that, without loss of generality, we can consider only
polynomial components.

\begin{cor}
Let $f$ be a polynomial with $f=g\circ h$, where $g,h$ are rational functions. Then there exists a
unit $u$ such that
\[g\circ u,\quad u^{-1}\circ h\]
are polynomials.
\end{cor}

Because of this, we only need to consider polynomial decomposition when our original function is a
polynomial. In the next section we will define and analyze the notion that will allow us to obtain
information about the decompositions of a polynomial.

\section{The fixing group of a polynomial}

In order to obtain information about the decompositions of a polynomial, we will introduce a
concept that comes directly from Galois Theory.

\begin{defn}
Let $f\in\K(x)$. The \emph{fixing group} for $f$ is
\[\Gamma_\K(f)=\left\{\frac{ax+b}{cx+d}: f\circ u=f\right\}<PSL(2,\K).\]

We will drop the subindex when there is no possibility of confusion about the ground field.
\end{defn}

This definition corresponds to one of the classical Galois applications between the intermediate
fields of an extension and the subgroups of its automorphism group, as the following diagram
shows:

\[\begin{array}{ccc}
 \K(x) & \longleftrightarrow & \{id\} \\
 & & \\
 | & & | \\
 & & \\
 \K(f) & \longrightarrow & \Gamma_\K(f) \\
 & & \\
 | & & | \\
 & & \\
 \K & \longleftrightarrow & PSL(2,\K) \\
\end{array}\]

\textbf{Remark.} As $\K(f)=\K(f')$ if and only if $f=u\circ f'$ for some unit $u$, we have that
the application $\K(f)\mapsto \Gamma_\K(f)$ is well--defined.

Next, we state several interesting properties of the fixing group, see \cite{Sev04} for details.

\begin{thm} $ $
\begin{description}
   \item[(i)] Given $f\in\K(x)\setminus\K$, $|\Gamma_\K(f)|$ divides $\deg\,f$. Moreover, for
any field $\K$ there is always a function $f\in\K(x)$ such that $1<|\Gamma_\K(f)|<\deg\,f$, for
example for $f=x^2\,(x-1)^2$ we have $\Gamma_\K(f)$=\{x,1-x\} for any $\K$.

   \item[(ii)] $|\Gamma_\K(f)|=\deg\,f\Rightarrow\K(f)\subseteq\K(x)$ is normal. Moreover, if
the extension $\K(f)\subseteq\K(x)$ is separable, then
\[\K(f)\subseteq\K(x)\mathrm{\ is\ normal}\quad\Rightarrow\quad|\Gamma_\K(f)|=\deg\,f.\]
\end{description}
\end{thm}

\section{Uniqueness of intermediate fields of the same degree}

First, we will define the class of polynomials on which we will work.

\begin{defn}
A polynomial $f\in\K[x]$ is \emph{tame} when $\mathrm{char}\,\K$ does not divide $\deg\,f$.
\end{defn}

The following result shows a nice property of tame polynomials.

\begin{thm}\label{unic-tame}
Let $f\in\K[x]$ be tame and $f=g_1\circ h_1=g_2\circ h_2$ such that $\deg\,h_1=\deg\,h_2$. Then
there exists a polynomial unit $u$ such that $h_1=u\circ h_2$.
\end{thm}

\begin{proof}
See \cite{Gut91}.
\end{proof}

Due to the equivalence given by Theorem \ref{equiv-dec-field}, the previous theorem is equivalent
to the uniqueness of intermediate fields of the same degree; that is, if $\K(h_1)$, $\K(h_2)$ are
fields between $K(f)$ and $\K(x)$ and $\deg\,h_1=\deg\,h_2$, then $\K(h_1)=\K(h_2)$.

This is not true if we drop the tameness hypothesis.

\begin{exmp}[\cite{Sch00}]
Let $\K=\F_2,\ \alpha^2-\alpha+1=0$ with $\alpha\in\F_4$. We have that
\[x^4+x^2+x=(x^2+\alpha x)^2+\alpha^{-1}(x^2+\alpha x).\]
\end{exmp}

In the case of rational functions, the result is also false.

\begin{exmp}[\cite{AGR95}]\label{contraej-extension-rac}
Let
\[f=\frac{\omega^3x^4-\omega^3x^3-8x-1}{2\omega^3x^4+\omega^3x^3-16x+1}\]
where $\omega$ is a non-real cubic root of unity in $\Q$. $f$ is indecomposable in $\Q(x)$.
However, $f=f_1\circ f_2$ where
\[f_1=\frac{x^2+(4-\omega)x-\omega}{2x^2+(8+\omega)x+\omega}\ ,\ f_2=\frac{x\omega(x\omega-2)}{x\omega+1}.\]
\end{exmp}

\begin{exmp}\label{ej-fixed-field1}
Let
\[f=x^2+\frac{1}{x^2}.\]

This function has two different decompositions of the same degree that are not equivalent:
\[f=\frac{1}{x}\circ x^2=(x^2-2)\circ\frac{1}{x}.\]
\end{exmp}

\section{Main result}

In relation to the existence of these fields we will discuss the generalization of the following
result:

\begin{thm}[\cite{BN00}]\label{beardon}
Let $p_1,\ldots,p_m\in\C[x]$ be non-constant and $k_1,\ldots,k_m,k$ be the orders of the groups
$\Gamma(p_1),\ldots,\Gamma(p_m),\Gamma(p_1\circ\cdots\circ p_m)$. Then $k$ divides $k_1\cdots
k_m$.
\end{thm}

We try to generalize this to polynomials with coefficients in any field. First we study the fixing
groups of these polynomials.

\begin{thm}
Let $\K$ be a field and $f\in\K[x]$ a tame polynomial. Then $\Gamma(f)$ is cyclic.
\end{thm}

\begin{proof}
First we prove that there are no elements of the form $x+b$ in $\Gamma(f)$ with $b\neq 0$. Let
$H=\{x+b: b\in\K, f(x+b)=f(x)\}<\Gamma(f)$.

If $\car\ \K=p>0$, any element $x+b\in H$ with $b\neq 0$ has order $p$, so the order of $H$ is
divisible by $p$. But the order of $H$ divides $\deg\,f$, therefore $H$ is a trivial group. If
$\car\ \K=0$, no elements of the form $x+b$ with $b\neq 0$ have finite order.

Let $a,b,c\in\K$ be such that $ax+b,\ ax+c\in\Gamma(f)$. Then $(ax+b)\circ(ax+c)^{-1}=x+c-b$, thus
$b=c$. Therefore, $B=\{a\in\K^*: \exists b\,|\,ax+b\in\Gamma(f)\}$, a subgroup of the
multiplicative group $\K^*$, has the same order as $\Gamma(f)$. But $B$ is cyclic, thus there
exists $a_0\in\K^*$ such that $B=\langle a_0\rangle$. Given the corresponding
$a_0x+b_0\in\Gamma(f)$, it is clear that every element of $\Gamma(f)$ is a power of it, therefore
$\Gamma(f)$ is cyclic.
\end{proof}

We can now generalize Theorem \ref{beardon} to the case of tame polynomials:

\begin{thm}\label{beardon-generaliz}
Let $\K$ be any field and $p_1,\ldots,p_m\in\K[x]$ be tame. Let $k_1,\ldots,k_m,k$ be the orders
of $\Gamma(p_1),\ldots,\Gamma(p_m),\Gamma(p_1\circ\cdots\circ p_m)$. Then $k$ divides $k_1\cdots
k_m$.
\end{thm}

\begin{proof}
It suffices to take $m=2$ and then use induction. Let $\gamma$ be a generator of the cyclic group
$\Gamma(p_1\circ p_2)$. As $p_1\circ p_2=(p_1\circ p_2)\circ\gamma=p_1\circ(p_2\circ\gamma)$, by
Theorem \ref{unic-tame} there exists a unit $\eta$ such that $p_2\circ\gamma=\eta\circ p_2$. Then
$p_1\circ p_2=p_1\circ p_2\circ\gamma=p_1\circ\eta\circ p_2$, therefore $p_1\circ\eta=p_1$. That
is, $\eta\in\Gamma(p_1)$ and its order $l_1$ divides $k_1$.

Also, $p_2\circ\gamma=\eta\circ p_2$ implies $p_2\circ\gamma^r=\eta^r\circ p_2$ for each integer
$r$. On one hand, taking $r=k$, we have $\eta^k=x$, thus $l_1$ divides $k$. On the other hand,
taking $r=l_1$ we have $\gamma^{l_1}\in\Gamma(p_2)$, that has order $l_2=k/l_1$. Therefore, as
$l_1|k_1,l_2|k_2$ y $l_1l_2=k$, we have $k|k_1k_2$.
\end{proof}

\section{Generalizations and future work}

In the rational case, as the uniqueness of fields of the same degree is not true in general (see
Examples \ref{contraej-extension-rac} and \ref{ej-fixed-field1}), we can think that this theorem
cannot be fully generalized. This is indeed the case, as the next example shows.

\begin{exmp}
Let
\[f={\frac{-1+33x^4+33x^8-x^{12}}{x^2-2x^6+x^{10}}}.\]
We have that
\[\Gamma_\C(f)=\left\{\pm\,x,\pm\,\frac{1}{x},\pm\,\frac{i(x+1)}{x-1},\pm\,\frac{i(x-1)}{x+1},\pm\,\frac{x+i}{x-i},\pm\,\frac{x-i}{x+i}\right\}.\]

The element $i(x+1)/(x-1)$ has order 3, and a function that is fixed by it is
\[h=\frac{x^3+(x-1)x+1-i}{(x-1)(x-i)}.\]

The field $\C(h)$ is not left invariant by every element of $\Gamma_\C(f)$, only by the three
elements in the subgroup (as they leave the generator fixed). For example it is easy to check that
\[h\circ(-x)\not\in\C(h).\]
\end{exmp}

Still, the following conjecture can be posed even if the proof is not valid in this case.

\begin{conj}
Theorem \ref{beardon-generaliz} is true for every rational function whose degree is not a multiple
of the characteristic of the field.
\end{conj}

A different direction that may allow for some generalization is given by the relation between the
degrees of the components for tame polynomials:

\begin{thm}[\cite{Sch00}]
If $\K(f)\cap\K(g)$ contains a polynomial $h$ such that $\deg\,h\not\equiv 0 \bmod \car\ \K$, then
\[[\K(f):\K(f)\cap\K(g)]=\frac{\mathrm{lcm}(\deg\,f,\deg\,g)}{\deg\,f},\]
\[[\K(f,g):\K(f)]=\frac{\deg\,f}{\gcd(\deg\,f,\deg\,g)}.\]
\end{thm}

Because of this, it is possible to consider that, as in Theorem \ref{beardon-generaliz}, not only
$k$ divides $k_1k_2$ but also $\gcd(k_1,k_2)$. The following trivial example shows that this is
not true in general:

\begin{exmp}
The function $x^4=x^2\circ x^2$ does not satisfy the above statement, since $4\nmid 2$.
\end{exmp}

In any case, we consider that it is of interest to study the classes of polynomials and rational
functions for which these statements hold.

\section*{Acknowledgement}

This work is partially supported by Spanish Ministry of Science grant MTM2004-07086.


\begin{thebibliography}{99}

\bibitem{AGR95} C. Alonso, J. Gutierrez, T. Recio, \emph{A rational function decomposition
algorithm by near-separated polynomials}. J. Symbolic Comput. 19 (1995), no. 6, 527--544.

\bibitem{BN00} A. F. Beardon, T. W. Ng, \emph{On Ritt's factorization of polynomials}. J. London
Math. Soc. (2) 62 (2000), no. 1, 127--138.

\bibitem{Cade} J. Cade, \emph{A new public-key cipher which allows signatures}. Proc.
2nd SIAM Conf on Appl. Linear Algebra, Raleigh NC (1985).

\bibitem{CFM} D. Casperson, D. Ford, J. McKay, \emph{An ideal decomposition Algorithm}.
J. Symbolic Comput. 21 (1996), no. 2, 133--137.

\bibitem{Gat90} J. von zur Gathen, \emph{Functional decomposition of polynomials: the tame case}.
J. Symbolic Comput. 9 (1990), no. 3, 281--299.

\bibitem{Gat91} J. von zur Gathen, \emph{Functional decomposition of polynomials: the wild case}.
J. Symbolic Comput. 10 (1990), no. 5, 437--452.

\bibitem{Gut91} J. Gutierrez, \emph{A polynomial decomposition algorithm over factorial domains}.
C. R. Math. Rep. Acad. Sci. Canada 13 (1991), no. 2-3, 81--86.

\bibitem{GRR} J. Gutierrez, T. Recio, C. Ruiz de Velasco, \emph{A polynomial decomposition
algorithm of almost quadratic complexity}. Proc. AAECC-6/88; L. N. Computer Science 357 (1989),
471--476.

\bibitem{GR00} J. Gutierrez, R. Rubio, \emph{CADECOM: Computer Algebra software for functional
DECOMposition}. Proceedings of the Second Workshop on Computer Algebra in Scientific Computing,
V.\ G. Ganzha, E.\ W. Mayr, E.\ V. Vorozhtsov, editors, Samarkand, Uzbekistan, Springer--Verlag
(2000), 233--248.

\bibitem{GRS02} J. Gutierrez, R. Rubio, D. Sevilla, \emph{On Multivariate Rational Function
Decomposition}. J. Symbolic Comput. 33 (2002), no. 5, 545--562.

\bibitem{GGR} J. Gutierrez, R. Rubio, J. von zur Gathen, \emph{Multivariate Polynomial
decomposition}. Applicable Algebra in Engineering, Communication and Computing, 14 (2003), no. 1,
11--31.

\bibitem{Sev04} J. Gutierrez, D. Sevilla, \emph{On Ritt's decomposition Theorem in the case of
finite fields}. Finite Fields and Their Applications 12 (2006), no. 3, 403--412.

\bibitem{KL89} D. Kozen, S. Landau, \emph{Polynomial decomposition algorithms}. J. Symbolic
Comput. 7 (1989), no. 5, 445--456.

\bibitem{Nag93} M. Nagata, \emph{Theory of Commutative Fields}. Translations of Mathematical
Monographs, Amer. Math. Soc. 125 (1993).

\bibitem{Net1885} E. Netto, \emph{\"Unber einen L\"uroth-Gordaschen Satz}. Math. Ann. 9 (1895), 310--318.

\bibitem{Sch82} A. Schinzel: \emph{Selected Topics on Polynomials}. Ann Arbor, University
Michigan Press, 1982.

\bibitem{Sch00} A. Schinzel, \emph{Polynomials with special regard to reducibility}. Cambridge
University Press, New York, 2000.

\bibitem{Zip} R. Zippel, \emph{Rational Function Decomposition}. Proc. ISSAC-91 (1991), ACM press, 1--6.

\end{thebibliography}
\end{document}